# Ultrahigh Q SOI Ring Resonator With Strip Waveguide

Bai Mou[1, 2], Yan Boxia[2,*], Qi Yan[2], Wang Yanwei[2], Han Zhe[2], Yuanyuan Fan[2],WangYu[2]

[1]*University of Chinese Academy of Science, Beijing, 10049, China*
[2]*Institute of Microelectronics of the Chinese Academy of Sciences, Beijing, 100029, China*
*Corresponding author: yanboxia@ime.ac.cn

**Abstract**

The design and experimental demonstration of a large-area, ultra-high-Q-factor and single-mode operation silicon ring resonator based on a strip silicon photonic waveguide are presented, which is suitable for photon gyroscopic application. The ring resonator consists of multimode straight strip waveguides (MMWGs), single-mode strip waveguide (SMWGs), and 90° Bezier–bends those connected by liner tapers. The SMWGs is used in the coupling region to achieve the mode coupling for fundamental mode and not exciting the higher-order mode, The MMWGs with 1.6μm width is designed to decrease the propagation loss, tapers make smooth straight-to-bend transition that reduce the inter-mode crosstalk at the interface of straight and bent waveguides, and the 90° Bezier–bend is designed to realize that the fundamental-mode maintained and propagated at the bent with ultralow loss and low intermode interference. Based on these designs, simulation results show that the resonance depth is 0.9979 while the resonator is in an undercoupled state with the coupling length L=2μm and coupling gap d=0.2μm. For the fabricated resonator, the ultrahigh Q factor of $1.54 \times 10^6$ is experimentally demonstrated, with the free spectrum range (FSR) of 0.036nm and the resonance depth is 0.9810.

**Keywords:** Ring Resonator; strip waveguide; high quality factor; photonic integrated

## 1. Introduction

In recent years, silicon photonics technology has been rapidly developed due to its high performance and compatibility with complementary metal oxide semiconductor (CMOS)[1][2], which has been widely used in optical communication systems. Compared with traditional MEMS and bulk fiber optic implementations, the ring resonator gyroscope based on silicon photonics has the advantages of compactness, light weight, high integration, easy mass production, and low loss. The basic working principle of the integrated optical gyroscope is optical Sagnac effect. The angular velocity of the gyroscope is determined by detecting the difference in the resonance frequency of the clockwise (CW) and counterclockwise (CCW) beams[3]. The superior performance of the integrated optical gyroscope largely depends on the high Q factor depends on low propagation losses and the large length of the resonant cavity[4]. However, the propagation loss caused by the material characteristics and etching process limits the Q factor of the silicon microring resonator[5].

In order to decrease the propagation loss and improve the Q factor of ring resonator, several approaches have been proposed for high Q resonators in SOI, e.g. race-track resonators[6-13], these silicon photonic resonators reported in the recent years that have a Q factor higher than $1\times10^6$. A Q factor as high as $2.2\times10^7$ was achieved for the resonators based on thick silicon ridge waveguides[6], which was fabricated by a non-standard SOI process, and standard SOI processes usually have high waveguide propagation loss that it more difficult to design ultrahigh Q factor ring resonators.

Some ring resonator with strip waveguides have been demonstrated on standard SOI processes platform, such as the multi-mode bend waveguide of resonator was based on modified Euler curves in Ref.[7], and in Ref.[8] that the ring cavity used multi-mode waveguides, their actual measured Q factors both were $1.3\times10^6$. Other high Q resonators used thin ridge waveguides on SOI have also been researched. And obtained a Q factor of $1.1\times10^6$ in Ref.[10] and in Ref.[11]. However, in these resonators the single-mode waveguide bend that based on the normal circular introduced notable scattering losses because they were narrow, and strong scattering happened at their sidewalls. Other silicon microring resonators with the Q factor of $1.7\times10^6$[12] and $2\times10^6$[13] was reported that based on the large-area multi-mode ridge waveguide, but such a silicon photonic ring resonator required a mode converter to connect the ridge waveguide to the strip waveguide and a thermally tunable coupler, which cause the devices more complex and larger in size.

In this letter, we present a large-area, ultrahigh Q factor silicon microring resonator based on the straight MMWGs, liner tapers with 100μm length and 90° single mode bent waveguides based on Bezier curve. The MMWGs is designed to decrease the propagation loss, the 90° Bezier-bent waveguide is previously proposed to reduce loss in single mode waveguide bends since Bezier curve is an adiabatic curve with gradual radius change[14].

Tapers can make smooth straight-to-bend transition that reduce the inter-mode crosstalk at the interface of straight and bent waveguide. The SMWGs is used in the coupling region to reduce the size of microring resonator. The designed high-Q factor resonator is realized with a simple standard single-etching process provided by a multiproject wafer (MPW) foundry. As a result, for the fabricated ring resonator, the ultrahigh Q factor of $1.54 \times 10^6$ is experimentally demonstrated, with the FSR of 0.036nm. It is expected that the present silicon photonic resonators will play a very important role in many applications, in particular for realizing the angular velocity gyroscope with high sensitivity.

## 2. Device and simulation

The device is designed based on the SOI platform with the thickness of 220nm for the silicon layer and 2μm for the buried silica layer. The schematic of the proposed ultrahigh-Q silicon microring resonator is shown in Fig. 1(a). The total length of the resonator is 17.3mm. Grating coupler (GC) is employed to couple the TE0 mode light from the access fiber to the strip waveguide. The resonator composed by MMWGs with 1.6μm width to obtain a low propagation loss and low intermode crosstalk, as well as SMWGs with width of 450nm which are used in the coupling region and waveguide bend.

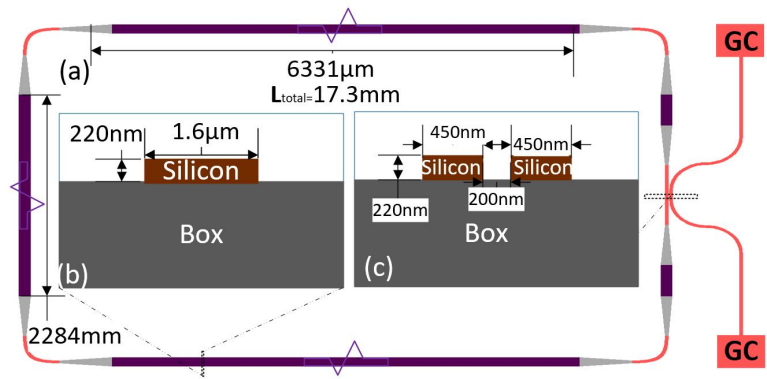

Fig. 1. (a) Schematic of the resonator formed by MMWGs waveguide, SMWGs, waveguide tapers, and GC. (b) Cross section of the coupling region. (c) Cross section of the MMWGs.

The coupled schematic of the proposed silicon microring resonator as show in Figure. 2(a). We designed a critical coupling directional coupler (DC) to obtain high Q factor, and then analyzed the optical power exchange between the waveguide and the resonator by consider the basic geometry. It consists of a lossless coupling between an optical waveguide and a ring resonator, the light wave E1 is input from the input terminal. Near-field-coupling is produced between the ring waveguide and bus waveguide due to the evanescent wave effect while the light wave passing through the straight waveguide[15]. Then part of the light field E3 enters the microring, after transmit a circle, the light wave E4 in the microring interferes with the light wave E1 which in the original bus waveguide. The resonance enhancement effect will be generated and the strong light field will be formed when the wavelength of light meets the basic resonance condition.

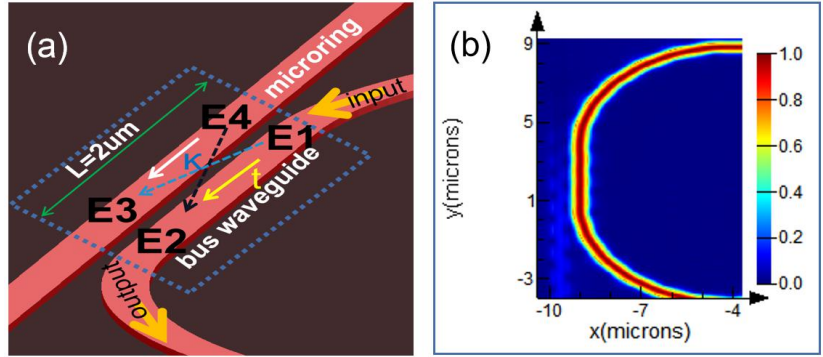

Fig. 2. (a) The description of bus waveguide couple to microring resonator. (b) Simulated field intensity $|E|^2$ for DC by FDTD when coupler length is 2μm.

After such multiple cycles, the light wave between the microring and the bus waveguide reached a balance. The light wave that satisfied the resonant condition confine to the microring, and the light wave what not satisfied the resonant condition will transmission out the bus waveguide[16]. Under the conditions that a single unidirectional mode of the resonator is excited and the coupling is lossless, we can describe the interaction by the matrix relation[17]

$$\begin{vmatrix} E_2 \\ E_3 \end{vmatrix} = \begin{vmatrix} t & \kappa \\ -\kappa^* & t^* \end{vmatrix} \begin{vmatrix} E_1 \\ E_4 \end{vmatrix} \qquad (1)$$

Where the complex mode amplitudes E1, E2, E3, E4, are normalized such that their squared magnitude corresponds to the modal power. The coupling coefficient and transmission coefficient are marked as κ and t respectively in the coupling region between the direct waveguide and the microring. Equation (2) is satisfied without considering the coupling loss.

$$|\kappa|^2 + |t|^2 = 1 \qquad (2)$$

Based on the mode coupling theory, when the light wave in the microring satisfies the basic resonance equation $2\pi R n_{eff}=m\lambda$, where $n_{eff}$ is the effective index of refraction, m is an integer, $\lambda$ is the wavelength, $2\pi R$ is the length of cavity. The transfer function of an optical waveguide resonator can be expressed as

$$T = t^2 = \frac{I_{output}}{I_{input}} = \left|\frac{E_2}{E_1}\right|^2 = \frac{t^2 + a^2 - 2at\cos\theta}{1 + t^2 a^2 - 2ta\cos\theta} \quad (3)$$

Where $a=\sqrt{10^{-\alpha L/10}}$ is the loss factor of light transmission in the resonant cavity for one cycle, and $\alpha$ is the transmission loss in an optical waveguide, L is the cavity length of the resonator, the phase $\theta$ of light wave in the microring depends on the integer m. Based on the assumption made on the expected waveguide loss, the condition for critical coupling has been imposed on the coupler transmission coefficient t:

$$t = a \quad (4)$$

The expected Q factor is obtained by determining the corresponding value of the power transmission coefficient and the coupling coefficient of the directional coupler. In this design, we assume a critically-coupled resonator of the total length as $L_{total}=17.3mm$, built entirely of MMWGs with $\alpha=0.3dB/cm$ propagation loss, the result display t=a=0.9319.

For the design of the DC, transmission coefficient t should be used to consider the coupling length of the DC, parameters of the directional coupler are designed as follow: the width of the bus waveguide and microring both are 450nm, the height of the silicon of 220nm. We considering the coupling gap from 190nm to 230nm between the waveguides and set coupling length L from 0μm to 5μm which the step size is 0.5μm. Simulated by the varFDTD method provided by lumerical MODE solutions[18].

As shown in the figure. 3(a). With the increase of coupling gap, the resonance depth decreases from 0.701 to 0.4612 while the transmission coefficient t gradually increases at the coupling length is 0um. The greater of the resonance depth, the better of the resonator coupling, as well as the limitation of manufacturing process error, Then choose the coupling gap of 200nm. As Figure. 3(b) show, the transmission coefficient t decreases gradually with the increase of coupling length while the coupling gap is 200nm, for the resonance depth, when L=2μm, there is a maximum value of $\rho$ =0.9979, it means that the coupling length and the coupling gap perfect matching in coupled region. So gap=200nm, L=2μm is ideal coupling length for this device in theory, where the transmission coefficient t is 0.9377 which slightly larger than the theoretical value (t=0.9319) and the coupling coefficient κ is 0.119(as Figure. 2(b)), known as undercoupling state.

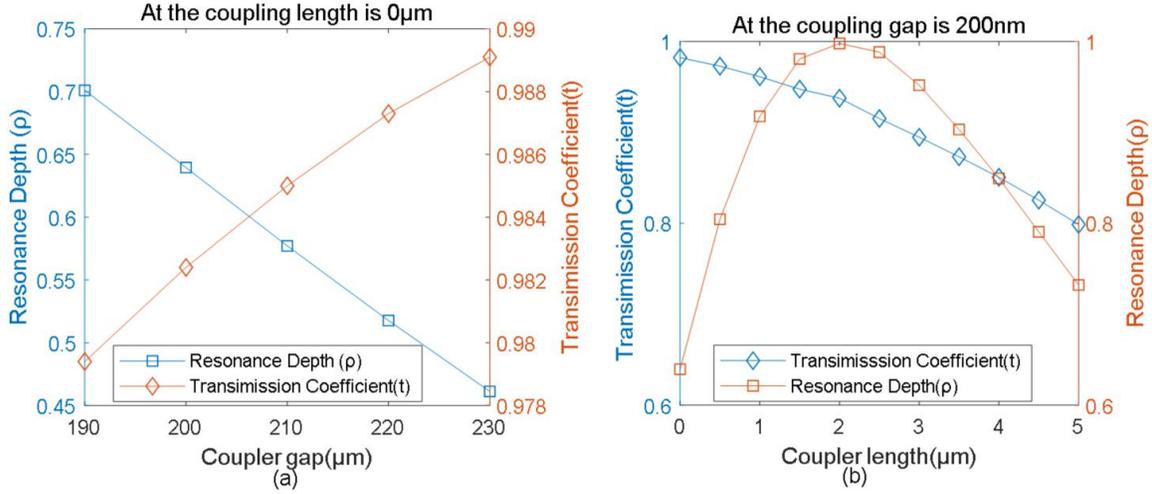

Fig.3. The transmission coefficient t and resonance depth ρ at: (a) different coupling gap with coupling length L=0μm, and (b) different coupling lengths with the coupling gap=200nm.

A 90° bent waveguide based on Bezier curve had been employed to reduce propagation loss in single-mode waveguide[14] and reduction of device size[19] in SOI waveguide devices. Bezier curve is an adiabatic curve with gradually change radius which defined by 4 points, the start point $P_0(x_0, y_0)$ and the end point $P_3(x_3, y_3)$. The other two midpoints describe the bend changes from beginning to end, they are defined as $P_1(x_1, y_1)$ and $P_2(x_2, y_2)$, and $x_0=0, y_0=0; x_1=R(1-B), y_1=0; x_2=R, y_2=B; x_3=R, y_3=R$. As defined by Ref. [20], so

$$B(t) = P_0 \times (1-t)^3 + 3 \times P_1 \times t \times (1-t)^2 + 3 \times P_2 \times t^2 \times (1-t) + P_3 \times t^3, \ t \in [0,1] \quad (5)$$

Where R is the radius of the curve and B is a unitless number that used to define how fast the radius of the curve changes. B is optimized to get the best crosstalk performance for the 90° bend with the three-dimensional finite difference time domain (3D FDTD) simulations. One work on Bezier curve focused on obtaining two-mode wavguide bends by set the R=20μm, B=0.15[19]. In this work, to avoid exciting higher-order modes at bend sections, we calculate the loss of the fundamental mode (TE0) and the mode excitation ratios of high-order modes over a wavelength band from 1500nm to 1600nm by MODE solutions[21], the TE0 mode is launched from the one end of wide waveguide. Through simulation, when R=5um, and B=0.2, we get the bend waveguide with single-mode operation, and the minimum loss of the fundamental mode is 0.0085dB after transmitted through the curved waveguide.

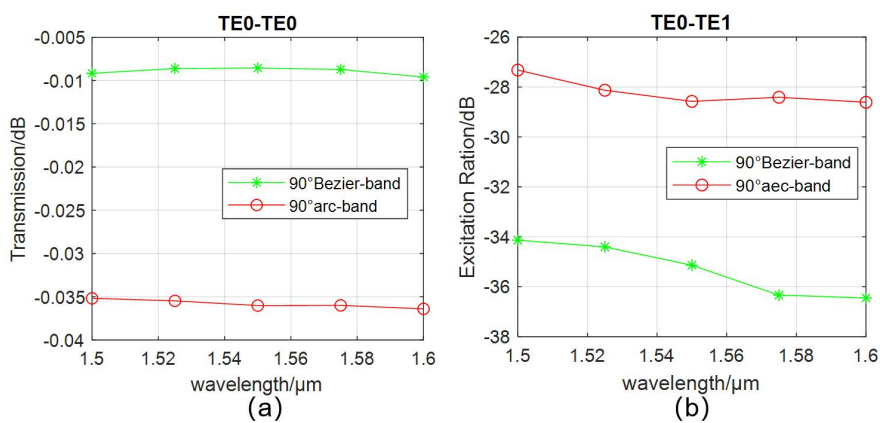

Fig.4. (a) The loss of the TE0 mode. (b) the MER of TE1 mode.

For comparison, we designed a normally circular 90° bent waveguide with the same radius, but its propagation loss is higher than 0.036dB, the obtained results were shown in Figure. 4(a). The MER for $TE_1$ mode of the 90° Bezier bend waveguide range from 34dB to 36.45dB and the MERs for the other high-order modes are very low that can be neglected. Meanwhile, the MER for $TE_1$ mode range from 27.5dB to 28.5dB in the 90° Bezier bend waveguide (shown in the Figure. 4(b)). It indicate that the Bezier curve design has a great deal of potential for the loss of the $TE_0$ mode and the MERs of high-order modes.

Taper is a basic device which is widely used in photonics field to transmit light between the waveguides with different widths which avoids modal conversion by narrowing the waveguide over a relatively long distance[22]. To ensure that only the fundamental mode of the MMWGs is propagated, in this work, we propose an efficient taper which have been employed to connect the SMWGs and the MMWGs. A linear waveguide taper that $W_0$=0.45μm is the minimum width of the single-mode waveguide and $W_1$=1.6μm is the maximum taper width, the length of the linear taper is $L_{taper}$=100μm. If the length of the taper is designed to be too short, the mode excitation ratios of high-order modes and the high loss of $TE_0$ mode is caused by the too small angle of the lines at the junction positions, on the contrary, it will introduce larger transmission loss[23].

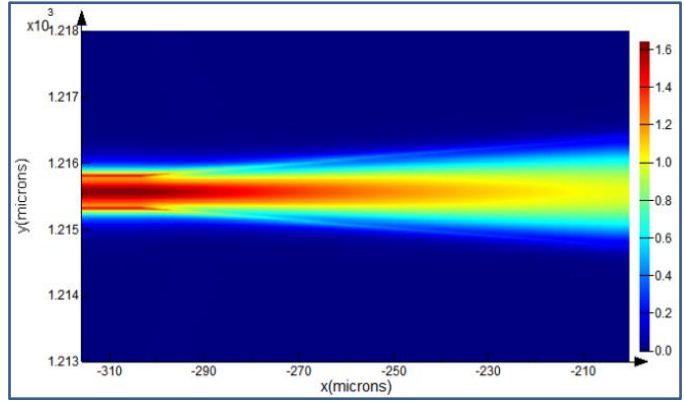

Fig.5. Simulated field intensity |E|² for the linear tapers at a wavelength of 1550nm.

Obtained the electric field |E|² of the linear tapper at a wavelength of 1550nm is shown in Figure. 5. during the transmission process and a transmission of 99.89%. The input light source was set with a fundamental mode and TE polarization which was launched from the 1.6μm wide MMWGs. This indication that the MERs of higher-order modes and the radiation of $TE_0$ mode can't be observed in this taper.

In short, the resonance depth is 0.9979 with the coupling length of 2μm and the coupling gap of 200nm by optimized in coupling region, that means the resonator in a undercoupled state, the 90° Bezier-band is designed to reduce the transmission loss of $TE_0$ and suppress the excitation higher-order modes. Furthermore, a linear taper with 100μm length is used to reduce radiation losses and to ensure only single-mode transmission at the output side. And assuming a critically-coupled resonator of same length that L=17.3mm, built entirely of MMWGs with α =0.3 dB/cm propagation loss, the expected Q factor would be $2.4 \times 10^6$ through simulation[7].

## 3. Results and discussion

The designed ultrahigh-Q resonators was fabricated by the multiproject wafer(MPW) foundry (Institute of Microelectronics, China) with the standard processes of deep UV lithography technologies and inductively coupled plasma dry etching. The Microscope and scanning electron microscopic (SEM) images of the fabricated silicon PIC and the devices are shown in Fig. 6(b) and 6(c). The color of pictures is light as a 1μm thick silica thin film was deposited on the top as the upper cladding.

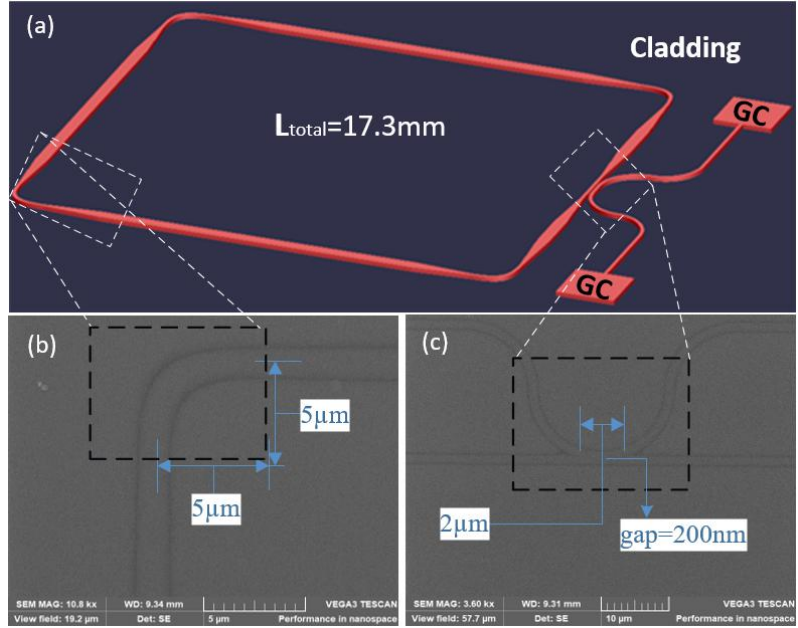

Fig.6. (a) Schematic of the resonator. (b)Top view of the 90° Bezier bend waveguide.(c)Top view of the directional coupler.

The measurement setup for characterizing the fabricated microring resonator as follow: a tunable laser source (Thorlabs C-BAND TUNABLE LASER) that is tunable from 1527.6nm to 1565.5nm and the step size of the tunable laser source is set to 20MHz, and the output signal of the MRR is detected by a photodetector (Newport 843-R Power Meter) associated with a data-acquisition computer. Then the relationship between real-time laser wavelength and optical power is obtained through LabVIEW. The measurement setup is adopted to measure the characteristic parameters of devices with coupling gap is 200nm and the coupling length are 2μm, 3μm, 4μm, 5μm, respectively.

Table 1 Shows the characteristic parameters of fabricated resonator

| Coupling length(μm) | Theoretical value(%) | Experiment value(%) | Q-Factor |
|---|---|---|---|
| L=2 | 99.79 | 98.102 | $1.54 \times 10^6$ |
| L=3 | 95.19 | 95.134 | $1.17 \times 10^6$ |
| L=4 | 84.95 | 90.788 | $6.9 \times 10^5$ |
| L=5 | 73.22 | 84.405 | $5.6 \times 10^5$ |

The result show in Table 1. With the increase of the coupling length, the Q factor of the resonator decrease from $1.54 \times 10^6$ to $5.6 \times 10^5$, and the experimental value ρ1 of resonance depth changes from 98.102 to 84.405 which is basically consistent with the theoretical value ρ0. It is verified experimentally that the resonator in an near critical-coupling state when coupling length L=2μm and gap=200nm. And when L is greater than 2μm, it is in the overcoupling state. In the following work, we will continue to explore the performance of couplers with coupling length between L=1μm and L=2μm.

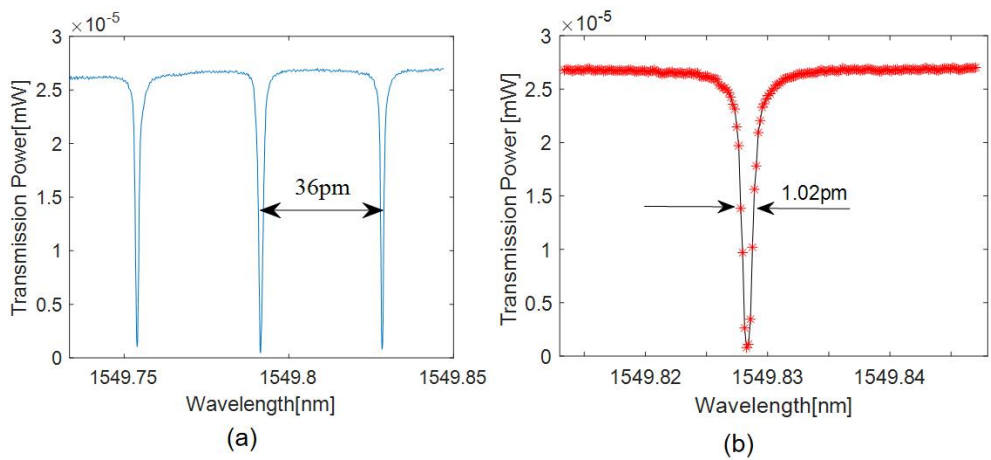

Fig.7. (a) Spectrum responses of a 17.3 mm long ring resonator, the FSR is 36 pm.
(b) Lorentzian curve fit of a resonance peak at 1549.828 nm, showing a FWHM of 1.02 pm.

Figure. 7(a) shows the spectrum responses obtained from the return signal of a circulator connecting the laser to the input GC. It can be seen that only the resonance peaks for the $TE_0$ mode were observed, while the higher-order modes are suppressed very well. Meanwhile the FSR is 36 pm. As Figure. 7(b) shows, the measured data shown by the red stars is fit using the theoretical lorentzian in Matlab (see the black solid curve). It can be seen that the FWHM of the resonance peak for the present resonator about 1.02pm, the Q factor can be calculated through the equation(6)[24], where $\lambda_c$ is the center wavelength of resonance, which indicates that a loaded ultrahigh-Q factor $Q_{load}$ of 1.54 × $10^6$ is obtained.

$$Q_{load} = \frac{\lambda_c}{\text{FWHM}} \qquad (6)$$

**Table 2** Large-Area, High-Q SOI ring resonators based on standard SOI process

| Ref. | Waveguide | Reff(μm) | Q-Factor | FSR(nm) | Loss(dB/cm) |
|---|---|---|---|---|---|
| [12] | ridge | 6000 | 1.7× $10^6$ | 0.017 | 0.085 |
| This work | strip | 2700 | 1.54× $10^6$ | 0.036 | 0.3 |

Table 2 shows a comparison of the Large-Area ring resonators reported in the recent years that have a Q-factor higher than 1×$10^6$ based on standard SOI processes platform.

In Ref.[12] the silicon microring resonators with a Q factor of $1.7\times10^6$ and the FSR is 0.017nm which based on the multi-mode ridge waveguide, however, the resonator length is as large as 37.7mm. Furthermore, such a silicon photonic ring resonator require a mode converter to connect the ridge waveguide and the strip waveguide, which introduces main loss. And the transmission loss of strip waveguide is much greater than the rib waveguide under the same manufacturing condition, so it more challenging to fabricate high Q factor resonators with strip waveguides. Compare to Ref.[12], we present the design and experimental demonstration of the ultrahigh-Q factor silicon microring resonator based on a strip silicon photonic waveguide. The high Q factor resonator is realized with a simple standard single-etching process provided by MPW foundry. As a result, For the fabricated ring resonator, the ultrahigh Q factor of $1.54 \times 10^6$ is achieved experimentally, with the FSR is 0.036 nm. To the best of our knowledge, this is the highest reported Q factor for a ring resonator with strip waveguides which fabricated using a standard SOI process, to date.

## 4. Conclusion

In conclusion, the superior performance of the integrated optical gyroscope largely depends on the large size of the optical cavity and high Q factor, the high Q factor depends on a large enclosed area, low propagation losses and the large length of the resonant cavity. We design and demonstrate a large-area, ultrahigh-Q factor and single-mode operation silicon ring resonator based on a strip silicon photonics waveguide, which consists of MMWGs, 90° Bezier–bends and SMMGs coupler. After fabricated the microring resonator, we obtain ultrahigh-Q factor of $1.54\times10^6$ and the free spectrum range of 0.036nm. The experimental demonstration shows that the proposed resonator can be successfully employed in high sensitivity ring resonator gyroscopes. And it is well suited for applications requiring high spectral selectivity, such as sensing, microwave photonics filters, lasers, and more.

**Acknowledgements**

This work is supported by the Key Deployed Project of Chinese Academy of Sciences (ZDRW-XH-2019-3) and Qingdao Optoelectronic Industry Think-Tank Joint Fund (GDZK-2019-4).